*Dense plasma irradiated platinum with improved spin Hall effect.*

Sachin Kumar[1], Sourabh Manna[2], John Rex Mohan[3], Utkarsh Shashank[3], Jospeh Vimal[4], Mayank Mishra[2], Surbhi Gupta[5], Hironori Asada[6], Yasuhiro Fukuma[3,7], Rajdeep Singh Rawat[2] and Rohit Medwal[1]*

[1]Department of Physics, Indian Institute of Technology Kanpur, Kanpur 208016, India

[2]Natural Sciences and Science Education, National Institute of Education, Nanyang Technological University 637616, Singapore

[3]Department of Physics and Information Technology, Faculty of Computer Science and Systems Engineering, Kyushu Institute of Technology, 680-4 Kawazu, Iizuka 820-8502, Japan

[4]The Ernst Ruska-Centre for Microscopy and Spectroscopy, Forschungszentrum Jülich, 52428, Germany

[5]Department of Physics, Motilal Nehru National Institute of Technology, Barrister Mullah Colony, Teliarganj, Prayagraj, Uttar Pradesh, 211004, India

[6]Department of Electronic Devices and Engineering, Graduate School of Science and Engineering, Yamaguchi University

[7]Research Center for Neuromorphic AI hardware, Kyushu Institute of Technology, Kitakyushu 808-0196, Japan

*Correspondence should be addressed to R.M.: rmedwal@iitk.ac.in

**Abstract:** The impurity incorporation in host high spin orbit coupling materials like platinum has shown improved charge to spin conversion by modifying the up-spin and down-spin electron's trajectories by bending or skewing them in opposite directions. This enables efficient generation, manipulation, and transport of spin currents. In this study, we irradiate the platinum with non-focus dense plasma to incorporate the oxygen ion species. We systematically analyze the spin Hall angle of the oxygen plasma irradiated Pt films using spin torque ferromagnetic resonance. Our results demonstrate a 2.4 times enhancement in the spin Hall effect after plasma treatment of Pt as compared to pristine Pt. This improvement is attributed to the introduction of disorder and defects in the Pt lattice, which enhances the spin-orbit coupling and leads to more efficient charge-to-spin conversion without breaking the spin orbit torque symmetries. Our findings offer a new method of dense plasma-based modification of material for the development of advanced spintronic devices based on Pt and other heavy metals.

## 1. Introduction

To enable practical spin-based logic and memory devices, achieving efficient charge-to-spin conversion [1,2] is crucial. In metallic systems, enhancing this charge to spin interconversion process has been a subject of significant research in spintronics [3–6]. Strategies such as utilizing heavy metals with strong spin-orbit coupling like platinum (Pt), tantalum (Ta), and tungsten (W) have been explored [7–12]. Additionally, alloying heavy metals with high atomic number metals like gold (Au) and palladium (Pd) allows for specific control of conversion efficiency [13–17], while incorporating adatoms, impurities and implantation in heavy metals with strong spin-orbit coupling can also enhance the spin-charge interconversion efficiency [18–22].

The localized implantation of diverse elements using ion beams has shown effective control over magnetic properties, driven by the need for fundamental physics understanding and technological advancements [2,5,23–25]. The versatility of ion-beam implantation has been explored extensively, leading to control over magnetization reversal mechanisms [26,27], exchange bias [28–31], perpendicular magnetic anisotropy in multilayers [32,33], spin-torque-based oscillator [34], spin-Hall nano oscillators [35–37] and bit-patterned magnetic recording media [38]. Recently, ion implantation has been employed to achieve enhanced spin-charge interconversion in materials [19,20,22]. Ion-implantation is a common and effective method which can be utilized to fabricate artificially engineered materials with large spin-orbit coupling for improved charge to spin conversion [39]. The enhancement of charge to spin conversion efficiency, defined by $\theta_{DL}$, arises from the interplay between the spin-orbit coupling of the 5d transition metal and the impurity-induced modifications such as nitrogen (N) in Pt, Ta, and W, oxygen (O) in Pt, Ta, and W and sulfur (S) in Pt [18–20,22].

The alternative technique to implant doping ions in material is dense plasma focus, (DPF), devices which can produce a wide range ion energy (typically few keV to MeV) in a single shot [40]. A DPF is a z-pinch device, which can produce hard and soft x-rays, neutrons, ions, and electrons. It consists of a capacitor which is charged to set high voltage and then discharged to a co-axial electrode assembly placed in a low-pressure vacuum chamber through a high voltage switch [40]. To minimize the effects of instabilities and achieve the uniform doping of the sample, the mode operation is switched to non-focus mode operation for this study [40,41]. Material synthesis like nitriding, oxidizing, and physical changes like amorphization by using DPF has been demonstrated [42,43].

The high-energy ions of the dense plasma can be used to create localized defects in targeted metals. In this work, we used the non-focus dense plasma to modify the platinum to improve charge to spin interconversion. The wide energy spectrum produced by the DPF helps in uniform incorporation of oxygen ion species across the thickness of the Pt. Thereafter, the oxygen treated Pt is subjected to the spin device fabrication and its characterisation for charge to spin conversion using spin torque ferromagnetic resonance technique. We observed a 2.4 times improved charge to spin conversion after oxygen plasma irradiation. Our findings provide an opportunity to understand the new method of incorporation of non-metallic elements into heavy metals (HM), through dense plasma-based techniques.

## 2. Experimental results and discussion:

To create the localized defects in a spin Hall material using the high-energy density plasma, multilayer stack of Pt (5 nm)/MgO (10 nm)/$Al_2O_3$ (10 nm) were deposited on Si/$SiO_2$ substrates at room temperature using an ultrahigh vacuum sputtering. The deposited stack was subjected to energetic plasma/ion processing of Pt using UNU-ICTP dense plasma device (DPF). The ion energy produced by DPF is unique and different from other ion sources/implanters. This provides plasma over a very wide energy range, utilized to implant uniformly at different penetration depths, thereby, creating a sample with a uniform ion dosage over its thickness. A schematic diagram of the DPF device with required electrical components is shown in Fig. 1 (a) and optical photograph of the device with electrode assembly is shown in Fig. 1(b). The DPF device consists of concentric electrode geometry which are connected to a charged high voltage capacitor though a swinging cascade spark gap switch. Once the spark gap switch is triggered, a symmetric discharge is set up between the central anode and the outer cathode at the closed end and the plasma sheath moves axially up towards the open end due to Lorentz force. At the top of the anode, the Lorentz force compresses the plasma in a tight column to form the pinch. This compression is responsible for producing high-energy high-fluence plasma/ions. These high energy species are used to create the defects in the desired materials. To avoid the damage of the material using fast plasma stream and highly energetic ions, we slow down the compression phase of plasma by operating DPF device under higher operating pressure known as slow-focus-mode. The slow-focus-mode will also result in moderate increase in pinch plasma volume which also help to increase the area of uniform processing. The electrical parameters of the DPF device operated in the focused mode and non-focused mode is shown in Fig. 1(c). The pinch formation in the non-focus mode is signified by the peak in the voltage signal. The delay in the pinch formation in the non-focused mode operation was achieved by increasing the operating pressure shown in Fig. 1(c).

Thereafter, X-ray photoelectron spectroscopy (XPS) technique was used to confirm the modification of platinum after plasma oxygen exposure. The XPS measurements were performed with Kratos AXIS Supra XPS that is equipped with an automated dual anode (Al/Ag Kα) X-ray monochromatic source. The survey scan and core level spectra of C (1s), and Pt (4f), were recorded. All the experimentally observed core level spectra were calibrated with the 285 eV binding energy of C (1s) as shown in the Fig. 1(d). Figure 1(e) shows two peaks associated with the Pt (4f) core level for pristine and oxygen plasma irradiated Pt [44]. The observed peaks are centered at 71.3 eV and 74.6 eV, corresponding precisely to the $4f_{7/2}$ and $4f_{5/2}$ states of pure metallic Pt, respectively. The energy difference between these two peaks measures 3.3 eV, as expected for pure metallic Pt. A shift in the binding energy was observed in oxygen plasma irradiated Pt, which are centered at 72.0 eV ($4f_{7/2}$) and 75.4 eV ($4f_{5/2}$) as shown in Fig. 1(e). The deviation in electronic states of platinum after oxygen plasma treatment confirm the incorporation of oxygen in Pt. This shift in the binding energy can be attributed to the chemical interaction between Pt and irradiated oxygen species. We also concluded that after irradiation of oxygen, the energy difference between two peaks is increased (3.4 eV) which leads to enhancement of spin orbit coupling in the $PtO_x$.

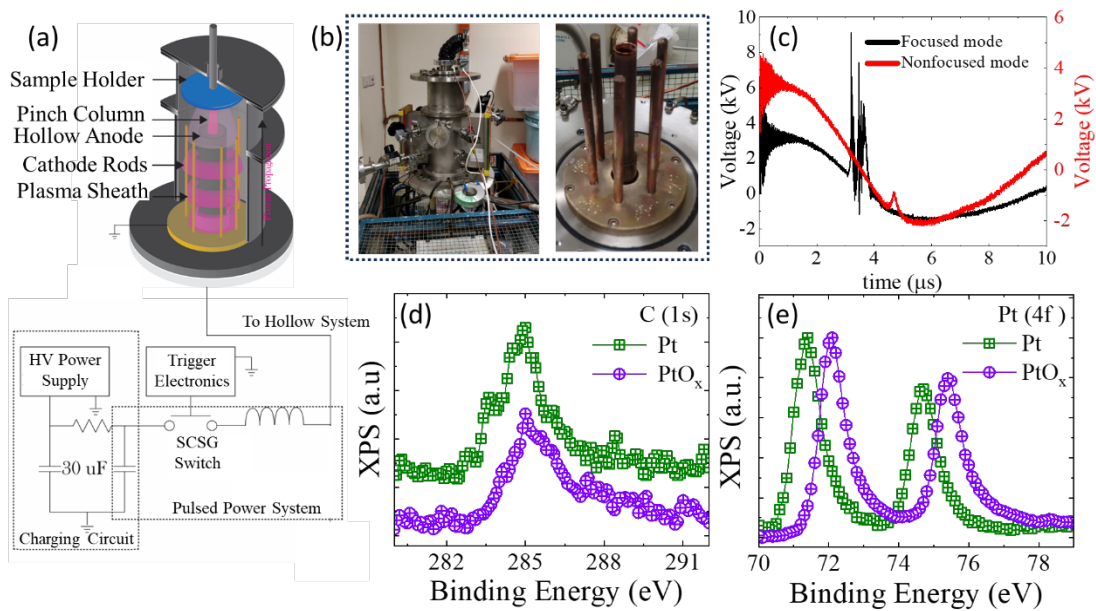

*Figure 1. (a). Schematic of the DPF device. (b) Optical photograph of the DPF device chamber used for the plasma treatment and electrode assembly. (c) Voltage profile showing the operation of DPF devices in non-focus mode (red) and focus mode (black). (d) XPS spectra of the C(1s) for the pristine and plasma treated Pt. (d) Core level spectra of the Pt (4f) for the pristine and plasma treated Platinum.*

After confirming oxygen plasma-based modification of platinum, both, modified platinum, and pristine platinum thin films stacks are subjected to ion milling to remove the 10 nm $Al_2O_3$ capping layer and 10 nm MgO end point detection layer followed by the deposition of 5 nm $Ni_{80}Fe_{20}$ ferromagnetic layer (hereafter referred as NiFe). The designed $Si/SiO_2/Pt/NiFe$ and $Si/SiO_2/PtO_x/NiFe$ bilayer are patterned into the ST-FMR devices.

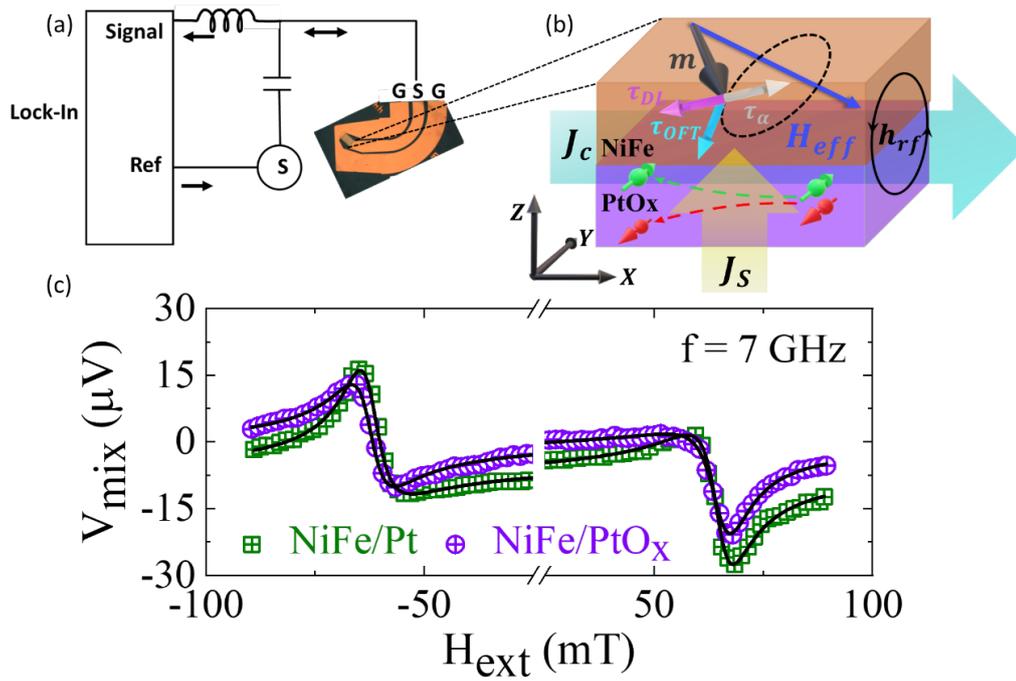

Figure 2. (a). Schematic of ST-FMR measurements scheme with optical photograph of micro-spin device used for the charge to spin conversion. (b) Illustration of charge to spin conversion phenomenon in the oxygen plasma treated platinum/NiFe heterostructure. (c) Spin torque ferromagnetic resonance (ST-FMR) spectra of the NiFe/Pt (square) and NiFe/PtOx (circle) heterostructures measured at 7 GHz excitation frequency.

To estimate the charge to spin conversion efficiency, $\theta_{DL}$, we initially performed the line-shape analysis, which uses the ratio of symmetric and antisymmetric component of ST-FMR spectra. The schematic diagram of the ST-FMR measurement setup is show in Fig. 2(a). In ST-FMR setup, when a microwave current, $I_{rf}$, passes in the longitudinal direction of the FM/HM interface, a transverse spin current, $J_s$, is generated by SHE, shown in Fig. 2(b). This spin current exerts an in-plane damping like torque (DLT), $\tau_{DL}$, on the magnetization vector of the FM layer. ST-FMR spectra was obtained with the input microwave power of $15\ dBm$. Simultaneously, an external dc magnetic field, $H_{ext}$, was applied at $\phi = 45°$, with respect to $I_{rf}$, in range of $\pm 90\ mT$. The $\phi$ is the angle of magnetization vector with respect to $I_{rf}$. The applied $I_{rf}$ induces a $rf$ Oersted field, $h_{rf}$, along the y direction, which exerts an Oersted field torque (OFT), $\tau_{OFT}$, on magnetization vector. At the resonance condition, both $\tau_{DL}$ and $\tau_{OFT}$ drive the magnetization vector of FM layer to sustained precession around the effective magnetic field, $H_{eff}$, as shown in Fig. 2(b). The torques produced by the $I_{rf}$ excite the magnetization vector, causing it to precess and deviate from equilibrium. This precession leads to a time-dependent alteration in the resistance of the bilayer of FM/HM, attributed to the anisotropic magnetoresistance (AMR) in the ferromagnet [20,45]. This change in resistance mixes with the $I_{rf}$ to generate a d.c. voltage, $V_{mix}$, across the bilayer. The in-plane DLT ( $\tau_\parallel = \tau_{DL}$) and out of plane OFT ( $\tau_\perp = \tau_{OFT}$) amplitudes defined in Fig. 2(b) contribute to the symmetric and antisymmetric parts of the $V_{mix}$ lineshape, respectively. Figure 2(c) shows the ST-FMR spectra of $V_{mix}$ measured at frequency, $f = 7\ GHz$ for both NiFe/Pt and NiFe/PtO$_x$ (for all spectra see supplementary S2) Individual components of torque are determined by fitting the $V_{mix}$ as a function of the $H_{ext}$ to a weighted sum of symmetric and antisymmetric Lorentzian functions.

$$V_{mix} = V_s F_{sym}(H_{ext}) + V_a F_{asym}(H_{ext}) \qquad (1)$$

Where, $F_{sym}(H_{ext}) = \frac{\Delta H^2}{[\Delta H^2 + (H_{ext} - H_{res})^2]}$, is symmetric Lorentzian function and $F_{asym}(H_{ext}) = \frac{\Delta H (H_{ext} - H_{res})}{[\Delta H^2 + (H_{ext} - H_{res})^2]}$ is the antisymmetric Lorentzian function with the resonant field, $H_{res}$, and linewidth, $\Delta H$. $V_s$ is weight factor of the symmetric Lorentzian component which is generated from the in-plane $\tau_{DL}$ because of injected spin current in FM layer and $V_a$ is the weight factor of the antisymmetric Lorentzian component which is generated by out-of-plane $\tau_{OFT}$ because of $h_{rf}$ induced from $I_{rf}$. Amplitudes of the Lorentzian function can be expressed in terms of torques [45]:

$$V_s = -\frac{I_{rf}}{2}\left(\frac{dR}{d\phi}\right)\frac{1}{\alpha\gamma(2H_{res} + 4\pi M_{eff})}\tau_\parallel \tag{2}$$

$$V_a = -\frac{I_{rf}}{2}\left(\frac{dR}{d\phi}\right)\frac{\sqrt{1 + 4\pi M_{eff}/H_{res}}}{\alpha\gamma(2H_{res} + 4\pi M_{eff})}\tau_\perp \tag{3}$$

Where $R$ is resistance of device, $dR/d\phi$ is anisotropic magnetoresistance, AMR, of the FM layer, $4\pi M_{eff}$ is the effective magnetization saturation, $\alpha$ is the Gilbert damping parameter and $\gamma$ is the gyromagnetic ratio. The parameters $V_s, V_a, H_{res}$ and $\Delta H$ are extracted by fitting the recorded ST-FMR spectra using equation (1). The recorded and fitted spectra STFMR spectra are shown in Fig. 2(c).

The symmetric and antisymmetric components of the $V_{mix}$ signal obtained at $f = 7\ GHz$ for NiFe/Pt and NiFe/PtO$_x$ are shown in Fig. 3(a) and (b), respectively. The increase in the amplitude of symmetric component of PtO$_x$ compared to Pt implies an increased $\tau_{DL}$ due to large injected $J_s$. The charge to spin conversion efficiency, $\theta_{DL}$, defined as $J_s/J_c$, is expressed as $\theta_{DL} = \frac{V_s}{V_a}\frac{e\mu_o M_s td}{\hbar}\sqrt{1 + \frac{4\pi M_{eff}}{H_{res}}}$ . where $e$ is the electronic charge, $\mu_o M_s$ is saturation magnetization, $t$ and $d$ are the thickness of NiFe layer and HM layer (here, Pt and PtO$_x$, respectively). The value of $\theta_{DL}$ estimated for PtO$_x$ is $0.154 \pm 0.005$, which is more than twice the value of $0.062 \pm 0.005$ estimated for Pt (see supplementary material S3).

The extracted values of $\Delta H$ are plotted as a function of $f$ as shown in Fig. 3 (c) and (d) for NiFe/Pt and NiFe/PtO$_x$, respectively. The effective Gilbert damping parameter, $\alpha_{eff}$, is estimated using a linear fitting function: $\Delta H = \Delta H_0 + \frac{2\pi\alpha_{eff}}{\gamma}f$, where $\Delta H_o$ is inhomogeneous linewidth, which is independent of $f$ and depends on sample quality. The value of $\alpha_{eff}$ for NiFe/PtO$_x$ is found to be $0.0201 \pm 0.001$, which is larger than that of $0.0174 \pm 0.001$ estimated for NiFe/Pt. The above calculation of $\alpha_{eff}$ is for positive $H_{ext}$ sweep (For negative field sweep, see supplementary material S4). Increased $\alpha_{eff}$ indicates the enhancement of spin injection into the PtO$_x$ layer. The $4\pi M_{eff}$ is estimated by fitting the Kittel equation: $f = \frac{\gamma}{2\pi}\sqrt{H_{res}(H_{res} + 4\pi M_{eff})}$ and the plot is shown in the inset of Fig. 3(c) and (d). The estimated value of $4\pi M_{eff}$ is 0.947 mT for NiFe/Pt and 0.944 mT for NiFe/PtO$_x$. The estimated values of $4\pi M_{eff}$ and $\alpha$ are in accordance with the values reported in literature [19]. The observed values of $4\pi M_{eff}$ suggests that oxygen irradiation has minimal impact on the on the sample roughness, thereby, magnetic properties of the NiFe alloy.

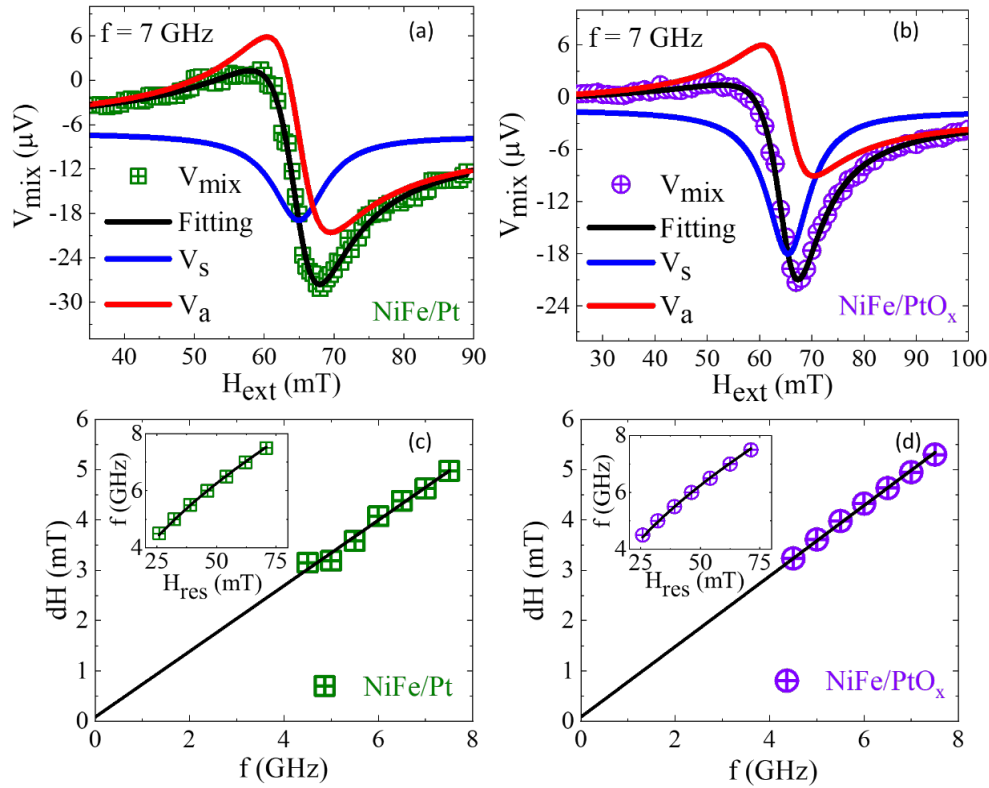

*Figure 3. The deconvolution of recorded voltage signal of ST-FMR signal ($V_{mix}$) into symmetric (blue) and antisymmetric (red) components for (a) NiFe/Pt (square) and (b) NiFe/PtO$_x$ (circle). (c) and (d) are $\Delta H$ vs $f$ with linear fit (solid lines) and inset shows the $f$ vs $H_{res}$ with Kittel equation fit (solid lines) for NiFe/Pt and NiFe/PtO$_x$, respectively.*

The symmetries of the device influence the strengths of current-induced torques, which vary with different angles of magnetization. Therefore, to investigate the symmetries of the torque in the pristine Pt and PtO$_x$, ST-FMR measurements were performed as a function of $\phi$ and shown in Fig. 4(a) and (b). In the case of a NiFe/Pt structure, the presence of a two-fold rotational symmetry involves that the sign of the spin orbit torque reverses when the magnetization is rotated within the in-plane by 180 degrees [45]. As a result, the $V_{mix}$ changes sign after $\phi = 180°$, while its magnitude remains same. The contribution to the expected angular dependence comes partly from the AMR as $sin2\phi$ dependence and partly from the in-plane spin-orbit torque as $cos\phi$ dependence. Consequently, in the absence of any other torque originated from the crystal symmetry, the overall angular dependence of $V_s$ and $V_a$ follows $sin2\phi cos\phi$. The weight factors of symmetric and antisymmetric components are observed to in be proportion to $sin2\phi cos\phi$ for both NiFe/Pt and NiFe/PtO$_x$ with no breaking of mirror and twofold symmetry for $\tau_{OFT}$ and $\tau_{DL}$. The normalized Fig. 4(a) and (b) shows the increase of symmetric weight factor of NiFe/PtO$_x$ as compared to NiFe/Pt, which implies that the increment in $\tau_{DL}$.

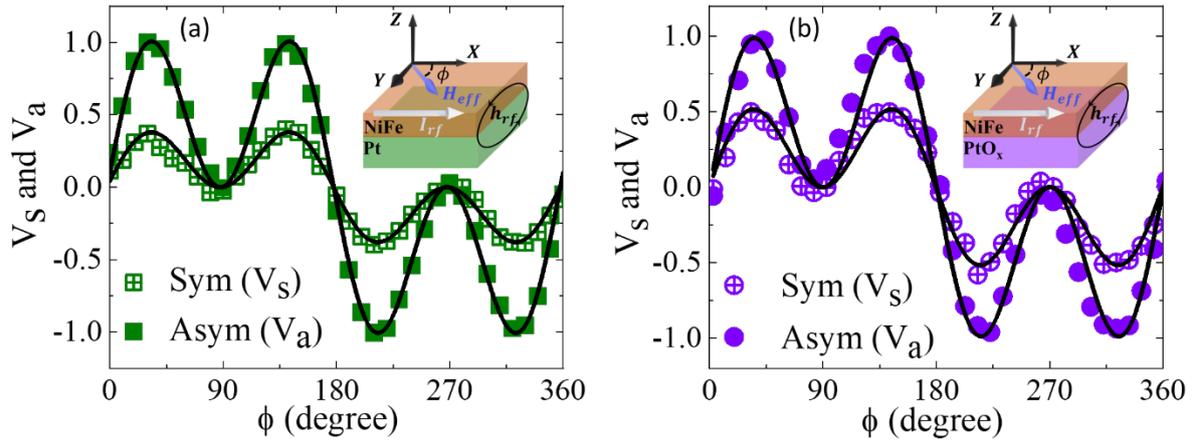

*Figure 4. Angular dependence of $V_s$ and $V_a$ components in ST-FMR spectra and corresponding fitted with to $\sin 2\phi \cos \phi$ for (a) NiFe/Pt and (b) NiFe/PtO$_x$*

**Conclusion:**

In conclusion, we have successfully demonstrated the non-focus dense plasma device as a versatile tool to incorporate the non-metallic ion species in the host materials. A significant enhancement (2.4 times) in the spin Hall angle of Pt after irradiation with oxygen ions using dense plasma focus techniques. The observed enhancement in the spin Hall effect can be attributed to the introduction of disorder and defects in the Pt lattice, which enhances the spin-orbit coupling. This, in turn, leads to more efficient generation and manipulation of spin currents. The ability to precisely control the implantation parameters and tailor the electronic and magnetic properties of the material through plasma irradiation provides opportunities for designing next-generation spintronic devices with enhanced performance and functionality.